# Topology Optimization for Materially Efficient Reinforced Concrete Design: Development, Fabrication, and Structural Evaluation

Jackson L. Jewett[1*], Josephine V. Carstensen[1]

## Abstract

The production of concrete generates roughly 8% of anthropogenic $CO_2$ globally, largely because of the massive quantities that are manufactured. New design methods must be developed and deployed to improve the material efficiency of reinforced concrete structures, and reduce concrete's carbon impact. This research uses topology optimization, a free-form structural optimization method, for improved structural design. Two topology optimization frameworks are developed specifically for reinforced concrete design and construction. The automated design algorithms are used to generate geometries for materially-efficient reinforced concrete beams, which are fabricated and tested to compare performance to conventional design. The optimized results exhibit ductile failure and reach loads 36%-42% higher than the conventional design with the same material consumption. Through comparison to analytical models, the observed potential for material reduction while maintaining today's performance requirements without adding structural depth is around 33%, indicating a viable path forward in reaching carbon neutrality of reinforced concrete construction.

[1]*Department of Civil and Environmental Engineering*
*Massachusetts Institute of Technology*
*Cambridge, United States*

*\*Corresponding author, jljewett@mit.edu*
*77 Massachusetts Ave, 1-170*
*Cambridge MA 02139*

# 1 Introduction

In the construction sector, cement production is an enormous source of anthropogenic greenhouse gas emissions. For every 1 kg of cement produced, 0.8 kg of carbon dioxide equivalent (co2e) emissions are released [4], with roughly 4 billion tons of cement being manufactured in 2024 alone [5]. Although cement is carbon-intensive to manufacture, it is typically about 10% of a normal concrete mix, giving concrete a lower co2e value than other structural materials like steel or even mass timber [6]. The concrete industry is thus such a large contributor of greenhouse gases in part because of the huge volumes of material that are produced. Concrete is the most used construction material in the world [7], [8], and it is estimated that global concrete consumption will at least double by 2060 with population growth, increased urbanization, and the need to rebuild infrastructure, [9], [10], [11]. Concrete also possesses some indisputable benefits: it is strong, durable, versatile, robust under natural hazards (e.g., fires and earthquakes), and low-cost [12], [13]. Therefore, extensive research has been devoted to limiting anthropogenic carbon associated with cement production, such as investigations in clinker replacement [12], [14], [15], fuel substitution, and improving cement plant thermal energy efficiency [16]. However, as outlined in the recently published Roadmap to Carbon Neutrality by the American Cement Association (ACA) [17] employing all of these known technologies can only achieve approximately half the reductions needed by 2050.

Significant efforts have therefor been devoted to identify embodied carbon reduction strategies for global construction, including minimizing the consumption of carbon-intensive materials through better design [18], [19], [20]. In this context, better design typically refers to structural solutions that minimize material usage using structural optimization tools [21], [22]. Because of its ubiquity in the field, optimization of reinforced concrete (RC) has received significant focus. These optimization methods must be tailored to account for the specific behavior of RC, particularly the interaction of the concrete and steel phases of the structural system.

Existing research in this field has demonstrated significant opportunities for material savings in RC construction [23], [24], [25]. Most works have focused exclusively on reducing the steel reinforcement in structures [26], [27], [28], [29], [30]. When including optimization of the concrete phase, previous studies have shown that using structural optimization can provide material reductions on the order of 10% when confining the design solutions to conventional prismatic structural shapes [23], [31], [32], [33]. However, these values fall short of the ACA's target to reduce global concrete use by 22% [17]. With new digital manufacturing technologies, such as CNC machining of formwork, flexible formwork, and concrete 3D printing [32], [34], [35], [36], [37], [38], [39], [40], [41], [42], [43], [44], the increased shaping opportunities of the concrete elements have received considerable attention [31], [32], [33]. Additional material savings are known to be possible when the geometric complexity increases [31], especially if combined with freeform structural optimization methods. In this context, topology optimization (TO) is a structural optimization approach with great potential to reduce concrete consumption. Because of its free-form nature, TO can often significantly outperform other computational automated design and optimization methods [45]. An example of a typical TO design is shown in Fig. 1.

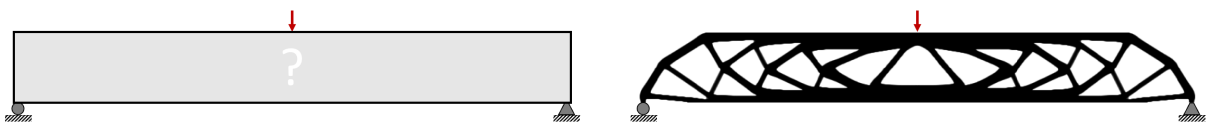

*Figure 1- A typical single-material topology optimization result is shown. On left, the design space is shown in gray, with load and boundary conditions for 3-point bending applied. On right, the materially-efficient beam design generated using topology optimization is shown. Material is placed in the black regions, while the white areas are void, and without material.*

TO works by modifying the stiffness, or "density," of elements in a finite element discretization of the design problem on a scale from 0 to 1. In a typical 2D TO problem, penalization schemes are used to promote density values of either 0 or 1. When using this approach, anywhere in the design space with a density of 1 is understood to have material, while locations with a value of 0 are void. This allows the optimizer to functionally turn on and off different parts of the structural system. The total amount of material in the system is limited as a constraint on the optimization. When the design problem is discretized with an adequately fine mesh, this can result in manufacturable, materially-efficient structures.

However, in its most basic formulation, TO controls the placement of a *single* material within a structural envelope, which means it cannot be directly used to design RC structures. Therefore, a tailored TO framework is needed to capture the design of RC structures that includes the distinctions between the manufacturing and behavioral characteristics of concrete and steel reinforcement [46].

There has been excellent research for over a decade in materially-specific TO for RC [47]. A popular approach for TO of RC is to generate strut-and-tie layouts, which has been done using a single continuum material [48], a truss system [49], or a combined continuum-truss system with force-dependent stiffness penalization [50] [51]. There are several examples of developing strut-and-tie models by using a continuum TO, then skeletonizing the results to generate a truss model [1], [52], [53]. RC elements have also been designed directly with TO using continuum optimization methods with different stiffnesses in compression and tension using one material [54] or two materials [55], [56]. The first papers using TO specifically for RC design used Drucker-Prager yield criterion to place material within a design space [57] [58], or a damage model with truss and continuum elements interacting [47], [59]. There has also been exciting work using cables and post-tensioning for concrete reinforcement [60], [61], [62], examples of which have been fabricated using 3D printing [63], [64].

This research builds on previous contributions by developing and comparing two new TO frameworks for RC that both capture the behavior of concrete and steel using a hybrid truss-continuum finite element discretization system with force-dependent stiffnesses. In the hybrid truss-continuum discretization [47], [50], [65], truss elements represent steel rebars acting in tension, while continuum elements represent cast concrete acting in compression. To enforce this behavior, an element has its stiffness reduced if it is acting in the opposite direction of what is desired [50], [66]. When this occurs, the element no longer significantly contributes to the design's behavior, making it materially inefficient and incentivizing material removal. The frameworks presented herein introduce a new layer of design freedom, as they also optimize the previously fixed locations of connections between the reinforcement and concrete phases of the design. To facilitate interaction between truss and continuum elements during the finite element analysis (FEA), this work uses the stiffness spreading method [67], [68].

Fig. 2 illustrates the difference between the two tailored TO approaches for RC design and construction that are formulated herein. In both cases, the geometry is initialized by defining a design space with applied loads and boundary conditions, which is discretized with both continuum and truss elements. To avoid relying on a predefined initial design layout, all continuum elements are initially given a uniform density value, and potential truss elements are defined in a ground structure [69]. From here, the two approaches differ. The first approach (referred to as Binary TO and shown in the purple box) uses a standard TO formulation that maximizes the structural stiffness and results in an extruded 2D design. This is similar to the current literature of TO for RC, which tends to result in beam structures resembling either a reinforced concrete truss, a hybrid plain concrete-steel truss, or a mix between the two. The purple box shows the direct algorithmic output when designing with Binary TO and a rendering of the fabrication-ready digital model. The second approach (Variable Thickness Sheet TO shown in the green box) explores the use of the Variable Thickness Sheet (VTS) TO applied to RC design with maximized stiffness. When VTS is applied to TO, intermediate density values are not penalized. Instead, all density values are interpreted as varying thicknesses out-of-plane. VTS has been used to design with isotropic materials since the 1970s [70] [56], [57], [58], [59] [75], with recent renewed interest [76], [77] as it tends to be computationally cheaper than conventional TO. VTS is an especially appealing method for concrete design because these designs tend to have a shallow out-of-plane thickness, and cast concrete can be easily shaped into complex forms when poured into molds. To discourage thicknesses that would be too small to carry load, this work encourages a minimum thickness [78].

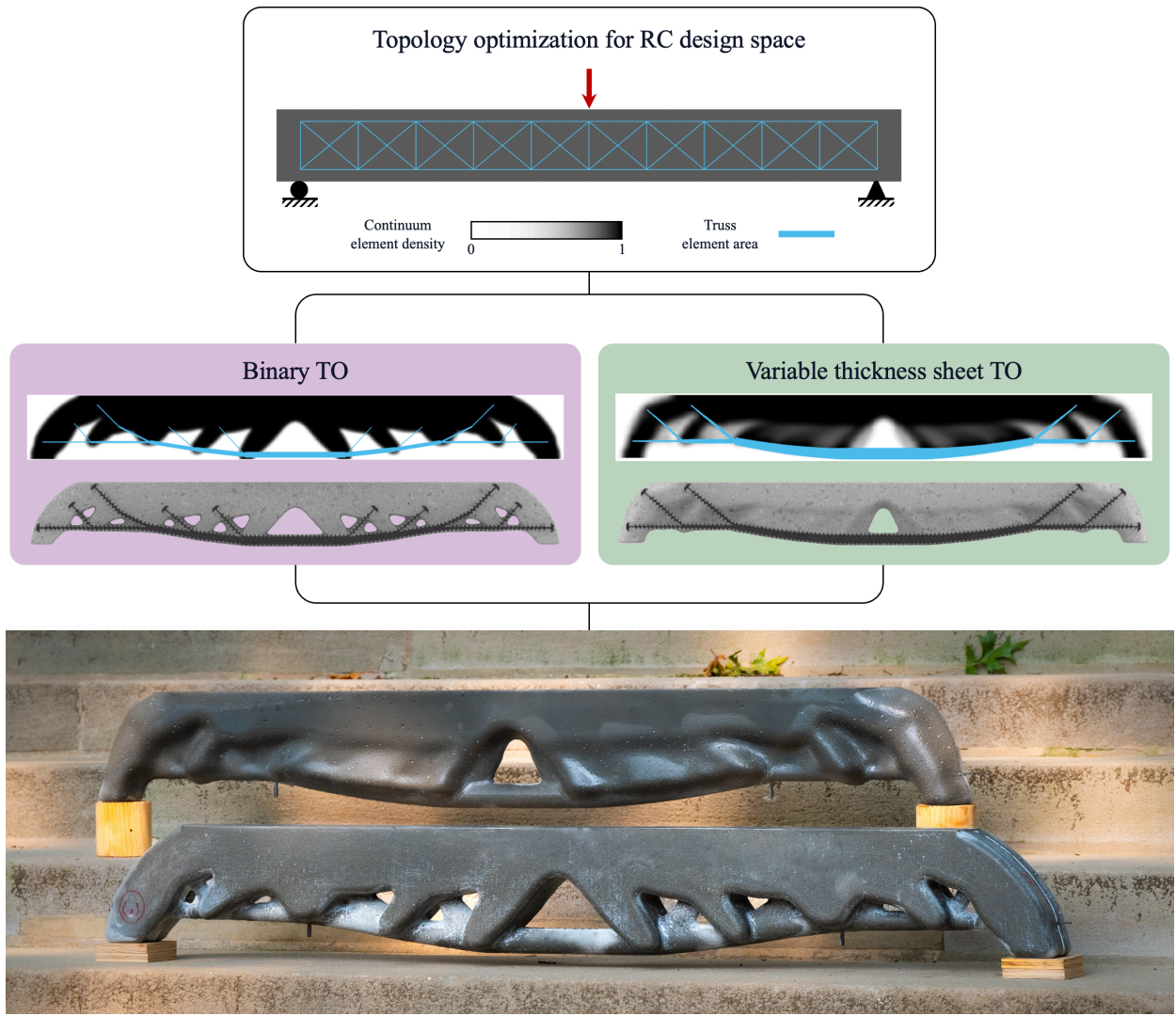

*Figure 2- This work develops and compares two new TO design methods that are tailored to RC design and construction. In both cases, the design process is initialized by the design engineer who defines a design envelope with applied loads and boundary conditions. The envelope then undergoes TO discretization using continuum and truss elements to represent the concrete and reinforcing steel phases of the design. The two approaches differ in that Binary TO encourages a 0-1 density design solution that results in an extruded 2D shape, whereas the Variable Thickness Sheet TO method interprets variable densities as out-of-plane thicknesses, which gives a more 3D, continuously shaped beam.*

Although computational gains point a way forward in limiting concrete consumption, constructability and experimental performance are crucial for actual deployment [79]. TO algorithms for RC have computationally proved to be economical and to save material [80], [81], but only a limited number of designs have been realized and experimentally validated [1], [3], [46], [49]. Nondestructive displacement testing in the elastic regime has been performed on TO 3D-printed concrete pedestrian bridges [63], [64]. Destructive testing that includes investigating failure modes and crack growth has focused on low-weight TO RC beams. Most have conducted static load testing, except for a single example of a fiber-reinforced concrete beam subjected to impact loading [82]. In the context of static testing of RC beams that mimic conventional reinforcement with steel rebars, previous works have demonstrated that 25-30% of the concrete mass can be removed while maintaining the same elastic performance as a standard by-code design [1], [83], [84], [85].

Further, a major challenge facing the use of TO in RC construction is the need to ensure ductile failure. A ductile structural response helps prevent brittle or catastrophic failure from occurring, which can be deadly in civil engineering projects. In the current TO for RC literature examples, many experimental investigations have reported failures induced by significant shear crack growth in the truss node regions, leading to catastrophic failure. Such behavior was observed in previous tests by the authors using TO RC [2], as well as several specimens in related work [3] [1]. As a result, ductility has recently become a major focus in the field, particularly in experimental testing [49].

To test the constructability and experimental performance of the two developed tailored TO design approaches, an experiment is conducted in which a Binary TO and Variable Sheet Thickness TO are designed for the same span, maximum depth, and maximum width. This work does not perform full-scale testing of RC structures, but limits the comparison to a prototype scale that has previously been shown to be effective for comparing TO RC designs [1], [2], [2], [3], [49], [83]. The same amounts of structural concrete and steel are allowed in both design cases. A Prismatic control specimen is also designed with the same material volumes and size constraints, using conventional methods, to allow comparison with the currently used design regime.

## 2 Methods

This work uses TO to design materially-efficient RC structures. The optimizer is able to control the placement of both plain concrete and steel elements within a design space to produce structural geometries. The density-based approach to TO is used, employing the Solid Intermediate Density Penalty (SIMP) method [86] to encourage manufacturable design solutions. The code used herein is built by modifying an open-source 2D TO continuum formulation [87], which is extended to include a 2D truss ground structure truss method [88] and force-dependent stiffness modification [66]. The Method of Moving Asymptotes (MMA) is used as the optimization engine for all designs herein [89]. All continuum elements referenced are linear square 4-node quadrilateral plane stress elements. The problem formulation is:

$$\begin{aligned}
\min_{x_c, x_p, x_t} \quad & c = \boldsymbol{F}^T \boldsymbol{d} \\
\text{s.t.} \quad & \boldsymbol{K}\boldsymbol{d} = \boldsymbol{F} \\
& \Sigma_{e \in \Omega} \left( \rho_{c,e} v_e \right) \leq V_{c,max} \\
& \Sigma_{f \in \Omega} \left( \rho_{t,f} A_f L_f \right) \leq V_{t,max} \\
& 0 \leq x_c \leq 1 \\
& \boldsymbol{P}_{min} \leq x_p \leq \boldsymbol{P}_{max} \\
& 0 \leq x_t \leq 1
\end{aligned} \quad (1)$$

Where $x_c$ are the design variables controlling the material distribution in the continuum element, $x_p$ are the design variables controlling the placement of the truss nodes, and $x_t$ are the design variables controlling the cross-sectional sizing of the truss elements. The objective function is the compliance, $c$, $\boldsymbol{F}$ is the finite element force vector, $\boldsymbol{d}$ is the displacement vector, and $\boldsymbol{K}$ is the global stiffness matrix. Continuum element $e$ has a volume $v_e$ and density of $\rho_{c,e}$, while truss element $f$ has an area of $A_f$, a length of $L_f$, and a density of $\rho_{t,f}$. $V_{c,max}$ and $V_{t,max}$ are the maximum volumes of continuum and truss material set by the user, respectively. $\boldsymbol{P}_{min}$ and $\boldsymbol{P}_{max}$ are the distances the truss nodes can be moved from their initial positions, which is also set by the user. Not all values in the $\boldsymbol{P}$ vectors need to be equal. In this work, the design variables $\boldsymbol{x}_c$ are filtered to define the density values, $\boldsymbol{\rho}_c$, which modify the element stiffnesses. The density filter [90] is used for the VTS optimization, and the Heaviside filter [91] is used for the Binary design. The truss elements are not filtered, so $x_{t,f} = \rho_{t,f}$.

For this work, three computational techniques are combined to approximate the behavior and construction of RC structures within the two new tailored TO design frameworks: (a) A force-dependent stiffness is enforced that allows truss elements (steel rebar reinforcement) to take tensile forces while the compressive internal forces in the design are guided to the continuum (cast concrete) elements, (b) the truss nodes can be moved within the design space, and (c) a Variable Thickness Sheet (VTS) method with a minimum thickness is employed. The force-dependent stiffness and stiffness spreading are parts of the general TO discretization and thus integrated in both the Binary TO and Variable Thickness Sheet TO design algorithms. Each of these techniques is explained below, followed by a section detailing the fabrication and experimental testing procedures used.

## 2.1 Force-dependent stiffness

In the hybrid truss-continuum TO used in this work, the continuum material represents concrete, and the truss elements represent steel. To ensure that continuum elements act only in compression and truss elements act only in tension, a force-dependent stiffness modification is enforced on the finite elements, as shown graphically in Fig. 3. This is similar to the work by Gaynor et al. [50]. The implementation of materials with force-dependent behaviors has been explored in many cases in the TO [54], [92], [93], [94], [95], [96]. The formulations outlined by Du et al. [97] are used herein, which have successfully been used for TO [66] as well as other works [98], [99], [100].

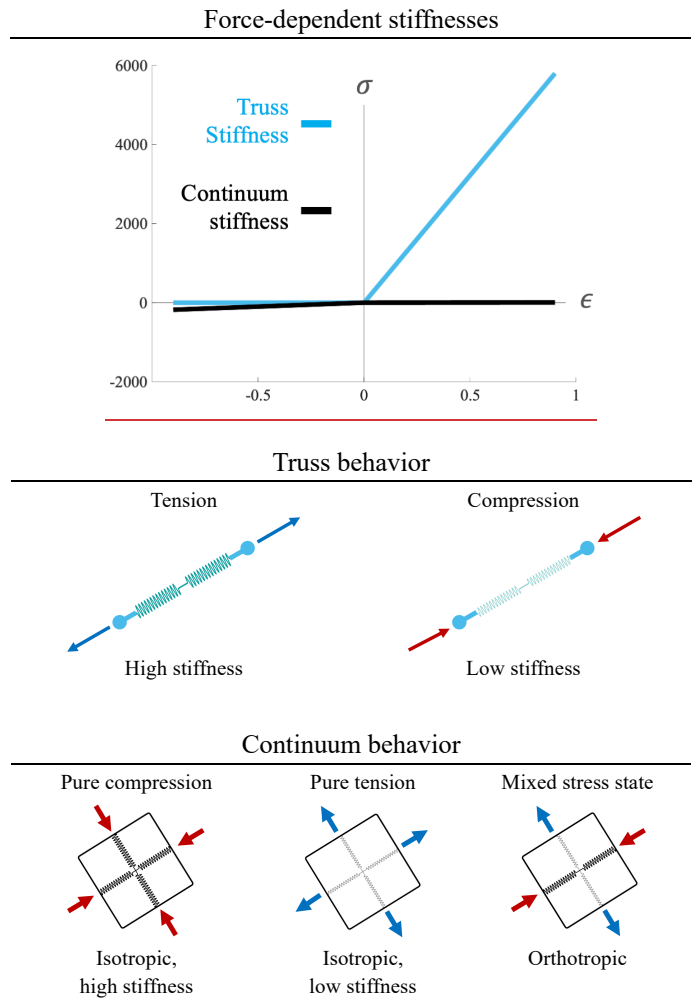

*Figure 3- On top, the stress-strain diagram for continuum elements (black) and truss elements (blue) are shown for both tension and compression stress conditions. Below, illustrations of how the stiffnesses of truss elements (middle) and continuum elements (bottom) are modified depending on their internal stresses. The dark springs represent a normal stiffness value, while the light springs illustrate a reduced stiffness in the element.*

The force-dependent stiffness modification is implemented as an initial loop that runs at the beginning of each optimization step. Within this loop, an FEA is run to calculate all elements' internal forces. The stiffnesses are then modified based on those values, with truss elements in compression and continuum elements in tension assigned low stiffness values. Changing the stiffnesses of the elements can change the behavior of the system, which in turn can change the directionality of some elements' internal forces. Therefore, this process must be repeated until the system converges. In this work, the loop is terminated when the value of the objective function changes by less than 0.1% relative to the previous calculation. After the loop converges, the stiffness matrix is set, and the outer TO algorithm continues as usual.

This stiffness modification is straightforward in the case of the truss elements, as they are 1-dimensional elements. In this work, the truss stiffness in tension is 5,800 and in compression it is 0.01, as illustrated in Fig. 3. These values are chosen because they allow the optimization to converge appropriately, and do not reflect the actual stiffness of structural steel.

The continuum elements are more complex. For these elements, an orthotropic material model is used, which has different stiffness values in its two principal directions. In this case, rather than assigning the elements a single elastic modulus and Poisson's ratio, $E$ and $v$, as is done in the isotropic case, an orthotropic material is used with $E_1$, $E_2$, $v_1$, and $v_2$, where:

$$\frac{E_1}{v_1} = \frac{E_2}{v_2} \qquad (2)$$

At each Gauss point of each finite element, a constitutive matrix is generated in global coordinates with the following formulation:

$$\boldsymbol{D}_{glo} = \frac{1}{1 - \nu_1 \nu_2} \begin{bmatrix} E_1 & \nu_1 E_2{}^1 & 0 \\ \nu_1 E_2 & E_2 & 0 \\ 0 & 0 & \frac{(1 - \nu_1 \nu_2)}{G} \end{bmatrix}, \qquad (3)$$

where

$$G = \frac{(1 + \nu_1)}{E_1} + \frac{(1 + \nu_2)}{E_2}. \qquad (4)$$

The values of $E_1$ and $E_2$ depend on the directionality of the principal stresses, as follows:

$$\text{if } \sigma_{1,2} < 0, \quad E_{1,2} = 180, \nu_{1,2} = 0.3$$

$$\text{if}, \sigma_{1,2} \geq 0, \quad E_{1,2} = 4.5, \nu_{1,2} = 0.0075.$$

This process is also illustrated in Fig. 3. In this work, the values of $E$ and $\nu$ are chosen to ensure the optimization converges, and do not refer to the real stiffness values of concrete. After the constitutive matrix has been assembled in global coordinates, it must be transformed by rotating the axes to align with the principal angle, $\theta$. This is done by multiplying $\boldsymbol{D}_{glo}$ by the transformation matrix $\boldsymbol{Q}$:

$$\boldsymbol{D} = \boldsymbol{Q}^T \boldsymbol{D}_{glo} \boldsymbol{Q}, \qquad (5)$$

where $\boldsymbol{Q}$ is defined as:

$$\boldsymbol{Q} = \begin{bmatrix} \cos^2 \theta & \sin^2 \theta & 2 \cos \theta \sin \theta \\ \sin^2 \theta & \cos^2 \theta & -2 \cos \theta \sin \theta \\ -\cos \theta \sin \theta & \cos \theta \sin \theta & \cos^2 \theta - \sin^2 \theta \end{bmatrix}. \qquad (6)$$

---

[1] Because $\nu_2$ is a function of $E_1$, $E_2$, and $\nu_1$, $C_{12}$ and $C_{21}$ can be written equivalently as $\nu_2 E_1$.

Each element's resulting $D$ matrix at each Gauss point is used the generate the element's stiffness matrices, $K_e$, such that:

$$K_e = \int B^T D B dV. \tag{7}$$

If the ratio of stiffness values in compression and tension is high, it is found that the system may not converge. In this work, a continuation method is used, in which the ratio begins at 0.3 and is reduced to 0.025 by increments of 0.025 until convergence is achieved.

## 2.2 Stiffness Spreading and Moving Truss Nodes

Previous works have shown that a high complexity of the reinforcement layout is not necessary to achieve significant mechanical performance improvements in terms of stiffness, crack growth control, and ultimate strength [2], [101], [102], [103]. A simple way to limit the complexity of the reinforcement layout is to restrict the complexity of the initial ground structure for the truss elements. However, this in turn means that the user predetermines the location of critical connection points in the reinforcement layout and forces connection between concrete and steel to happen at certain fixed locations. To expand the freeform nature while still having a low complexity of the initial truss ground structure, this work assigns design variables $x_p$ to the locations of the truss nodes. This allows the trusses to be moved during the design process by the optimizer, making the initial layout of the trusses less influential on the overall performance of the design.

The formulation for the moving truss nodes follows the procedures outlined by Achtziger [108] and, similarly, by Xia et al. [104], [105]. For each node $v_i$, the influence on the performance of the structure is the sum of the influences of each individual truss element attached to that node.

The sensitivity of this process can be formulated as follows:

$$\frac{\partial c}{\partial x_i^p} = -\sum_{e \in E(v_i)} \boldsymbol{u}_e^T \frac{\partial \boldsymbol{K}_e}{\partial x_i^p} \boldsymbol{u}_e, \tag{8}$$

$$\frac{\partial \boldsymbol{K}_e}{\partial x_i^p} = \frac{\partial \boldsymbol{T}_e^T}{\partial x_i^p} \bar{\boldsymbol{K}}_e \boldsymbol{T}_e + \boldsymbol{T}_e^T \frac{\partial \bar{\boldsymbol{K}}_e}{\partial x_i^p} \boldsymbol{T}_e + \boldsymbol{T}_e^T \bar{\boldsymbol{K}}_e \frac{\partial \boldsymbol{T}_e}{\partial x_i^p}. \tag{9}$$

Where $\boldsymbol{K}_e$ is the truss stiffness matrix in global coordinates, $\bar{\boldsymbol{K}}_e$ is the stiffness matrix in local coordinates, and $\boldsymbol{T}_e$ is the transformation matrix, which are defined below as:

$$\boldsymbol{T}_e = \begin{bmatrix} C & S & 0 & 0 \\ -S & C & 0 & 0 \\ 0 & 0 & C & S \\ 0 & 0 & -S & C \end{bmatrix}, \tag{10}$$

$$C = \frac{(x_2 - x_1)}{\sqrt{(x_2 - x_1)^2 + (y_2 - y_1)^2}}, \tag{11}$$

$$S = \frac{(y_2 - y_1)}{\sqrt{(x_2 - x_1)^2 + (y_2 - y_1)^2}}, \tag{12}$$

$$\bar{\boldsymbol{K}}_e = \frac{E_e A_e}{\sqrt{(x_2 - x_1)^2 + (y_2 - y_1)^2}} \begin{bmatrix} 1 & 0 & -1 & 0 \\ 0 & 0 & 0 & 0 \\ -1 & 0 & 1 & 0 \\ 0 & 0 & 0 & 0 \end{bmatrix}. \tag{13}$$

Here $x_1, x_2, y_1$, and $y_2$ are the coordinates of the truss nodes. $E_e$ is the elastic modulus of truss $e$, and $A_e$ is the area.

A complication when moving the truss nodes is that the stiffness of the truss elements must be properly transferred to the continuum elements in the system. In a typical FEA, the nodes must be identically shared for forces to transfer. However, in the formulation used herein, truss nodes are moved in a smooth manner to ensure the process is differentiable for gradient calculations. Because the nodes are not moved by discrete increments, it is impossible to ensure that truss nodes are perfectly coincident with the continuum nodes in the system. To address this challenge, this research uses the Stiffness Spreading Method (SSM). In SSM, the stiffness of a truss element is not transferred at a single node, but rather, it is distributed to several continuum nodes within a region of influence defined by the user, as shown schematically in Fig. 4. SSM has previously been used by Zegard and Paulino [68] to optimize the layout of truss elements within a continuum, as well as Wei et al. [67].

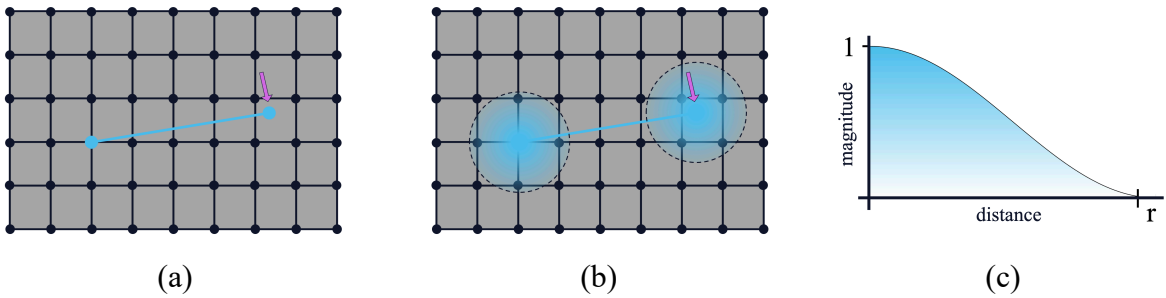

*Figure 4- In (a), a truss node has been moved away from the nodes of the continuum elements, and can no longer transfer stiffness appropriately. In (b), the stiffness spreading method is used, so that the stiffness can be transferred between truss and continuum element seven when the truss node is moved. In (c), the stiffness distribution function is shown as a function of distance between nodes.*

In principle, SSM is achieved by modifying the $N$ matrix, which is typically used to ensure that truss-continuum structural systems are compatible in FEA by distributing stiffness values from the truss elements to the degrees of freedom of the continuum elements. $N$ maps the truss element degrees of freedom (DoFs), $u$, to the continuum DoFs, $u_c$, such that:

$$u = Nu_c \qquad (14)$$

The $N$ matrix can then be used to modify the truss stiffness matrix to be compatible with the continuum stiffness matrix with the following formulation [68]:

$$\begin{aligned} u^T K_e u &= u_c^T K_e^+ u_c, \\ (Nu_c)^T K_e (Nu_c) &= u_c^T K_e^+ u_c, \end{aligned} \qquad (15)$$

$$u_c^T(N^T K_e N)u_c = u_c^T K_e^+ u_c,$$
$$N^T K_e N = K_e^+,$$

where $K_e$ is the stiffness matrix of the element, and $K_e^+$ is element stiffness matrix that has had its stiffness values assigned to the degrees of freedom corresponding to the continuum stiffness matrix.

$N$ is typically a matrix containing only 1s and 0s. In SSM, this process is modified to use $\widetilde{N}_e$, which is a matrix in which the values are defined by the distance of the truss nodes from nodes of the continuum elements. Each truss element has its own corresponding $\widetilde{N}_e$ matrix that is of dimension 4x$N_c$, where $N_c$ is the number of DoFs in the continuum system. The new formulation is then:

$$\widetilde{K}_e^+ = \widetilde{N}_e K_e \widetilde{N}_e. \tag{16}$$

Here $\widetilde{K}_e^+$ is truss stiffness matrix with SSM implemented. In this work, the stiffness distribution function is set as a modified sine function, which is visualized in Fig. 4c. The values of the matrix, $\widetilde{N}_{i,j}$, must be normalized so that the sum of all magnitudes for each truss node is equal to 1. This ensures SSM does not increase or decrease the stiffness contribution of the element. For this reason, initial values are calculated as a function of distance, $\breve{N}_{i,j}$, which are then normalized to define $\widetilde{N}_{i,j}$.

$$\breve{N}_{i,j} = \begin{cases} \frac{1}{2}\cos\left(\frac{\pi d}{r}\right) + 0.5 & \text{if } d \leq r, \\ 0 & \text{otherwise}. \end{cases} \tag{17}$$

where

$$d = \sqrt{(x_j - x_i)^2 + (y_j - y_i)^2}, \tag{18}$$

$$\widetilde{N}_{i,j} = \frac{\breve{N}_{i,j}}{\sum_{a=1}^{ndof} \breve{N}_{i,a}}. \tag{19}$$

In Eq.s (17-19), $d$ is the distance between the truss node and the continuum node, $r$ is the distribution radius selected by the user, $x_i$ and $y_i$ are the coordinates of the truss node and $x_j$ and $y_j$ are the coordinates of the continuum element.

## 2.3 Variable Thickness Sheet TO

Typically in TO, penalization schemes such as SIMP [106] are used to promote approximately discrete solutions in 2D and 3D. However, this may not be an appropriate constraint for all 2D design cases. Sigmund et al. [107] suggest that, from a theoretical perspective, forcing 0-1 outputs is a significant limitation on optimality, and leads to poor-performing geometries compared to the less restricted case allowing intermediate densities. An alternative suggestion is to allow continuous variable outputs in the final solution, and linearly interpolate those values as thicknesses. In this way, a value of 0 would have no material, a value of 1 would have a maximum chosen thickness, and values in-between would be linear interpolations. This is referred to as the Variable Thickness Sheet (VTS) method [45], and is illustrated in Fig. 5. The optimality of VTS was understood as early as 1973 by Rossow and Taylor [70], who are also referenced by Bendsøe and Kikuchi [108]. The VTS problem has been studied for decades for problems of in-plane forces [71], [72], [73], [74] and out-of-plane [75]. The strategy has recently regained interest in so-called 2.5D optimization [76], [77].

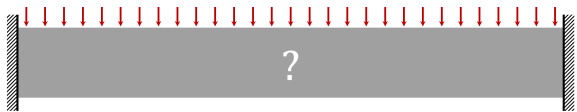

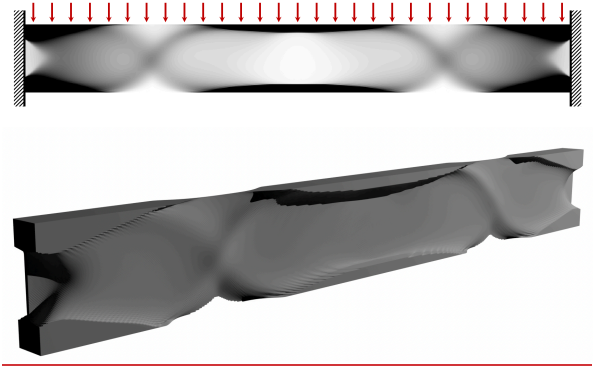

*Figure 5- Illustration of the design process using variable thickness sheet optimization. For a given design space (top), element density values are permitted to take intermediate values between 0 and 1 without penalization (middle). These densities are then linearly interpreted as thicknesses within the design space to generate the finale 3D geometry (bottom).*

To the authors' best knowledge, VTS TO has not previously been developed for RC design, despite its appeal, as cast concrete can be easily shaped into complex forms by pouring it into molds. However, in this work, a pure VTS optimization is not practical. It has been observed that VTS optimization often produces large zones of low-stiffness, but non-zero, densities [78]. Any thin concrete sheets designed in accordance with these low-density regions would crumble as they are removed from the mold, or fail at low stress values. In this case, it is more desirable to have a minimum thickness enforced on the material, with varying thicknesses admissible above that threshold. For this reason, a modified stiffness penalization function is used, which reduces the element stiffness for low values, but maintains a nearly linear relationship above a chosen threshold. This is achieved with the following formulation:

$$y = \frac{x}{1 + e^{(-x+m)C}}. \tag{20}$$

This relation is similar to the formulation used in [109]. The technique is used on both concrete and steel elements in this work.

A challenge that arises with the Variable Thickness Sheet TO for RC developed herein is that the optimizer masses continuum material around the truss nodes, rather than distributing it along the elements. This is because stiffness cannot be transferred along the length of the trusses, as they cannot take bending forces. This problem can be solved by making the truss elements very short, but this solution also poses problems, because the optimizer has difficulty moving the nodes effectively when they are very close together. The solution used herein is to begin the optimization with large truss elements and slowly divide them in half over the course of the optimization. The VTS truss nodes are allowed to move further in the $y$ direction than in $x$ to ensure that the nodes do not move on top of each other, as these divisions make the truss elements progressively smaller. This approach allows proper placement of the elements, and also gives enough nodes for the continuum material to distribute along.

## 2.4 Fabrication and Experimental Testing

To fabricate the RC specimens, reusable two-part molds are made. The pieces are cut from solid polyethylene blocks using a CNC mill with a 1.9 cm (3/4") ball end mill using the "Steep and Shallow" cutting routine with 1.27 mm (0.05") stepover and 1.9 mm (0.075") stepdown. The two halves are fixed together and filled from the top with concrete or mortar. Small registers are cut into the molds that align with protrusions included on the reinforcing steel, which allows them to align and thus ensures the steel is placed properly when set into the molds. The construction process is shown conceptually in Fig. 6, along with a photo of one of the fabricated molds.

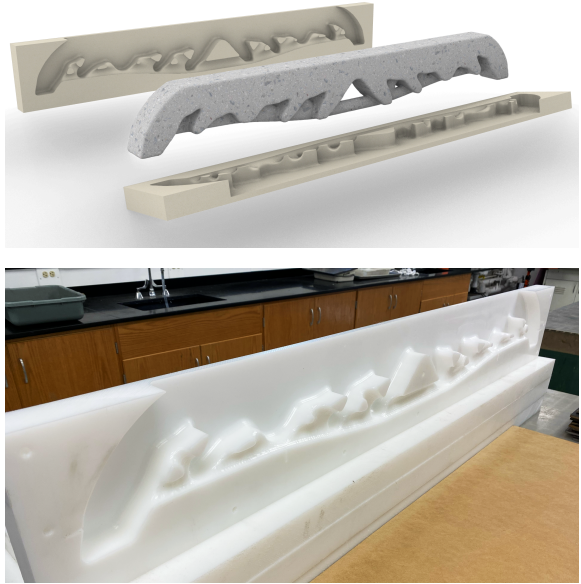

*Figure 6- On top, a rendering of the Binary TO mold designs, showing how two symmetric halves can fit together to be filled with concrete from above. Below, a photo of one of the fabricated Binary TO molds.*

The steel is fabricated by cutting the desired geometries from 6.53 mm (0.25") thick steel plates rated for 248.2 MPa (36 ksi) yielding, although tensile tests of dogbone samples cut from these same plates revealed the strength to be roughly 372.3 MPa (54 ksi). An OMAX abrasive waterjet machine with 80 grit Garnet abrasive was used to cut the steel. A cut quality of 2 was used. Small non-structural protrusions are added to the steel to ensure alignment within the molds. The reinforcing steel was cut from plates as three separate pieces and welded together. Photos of the fabricated steel are shown below in Fig. 8.

Because the fabricated beams are relatively small, mortar is used instead of concrete, as large aggregate may not fit into the small portions of the molds. This approach has been used in other TO for RC research [1]. Additionally, a low viscosity mortar is desired for pouring, so the Quikrete 3D Printing Mix is used. The mix is rated for 34.5 MPa (5 ksi) compressive strength with a 28-day cure. Three concrete batches were mixed, one for each beam sample, and 3 cylinder samples were taken from 2 of them for strength tests, which showed average compressive strengths of 57.2 MPa (8.3 ksi) and 56.5 MPa (8.2 ksi). Using a 15.2 cm (6" tall) cone, the concrete mixes had an average slump of roughly 14 cm (5.5"), which corresponds to the concrete acting as a liquid and puddling.

During fabrication, the steel is placed into one half of the mold. The other half of the mold is then set in place and secured with steel bolts. After the mortar is poured to fill the mold, a vibration table is used so the mortar settles evenly. After an initial 24-hour cure, the concrete is removed and placed into a water bath to complete the 28-day cure. One reusable mold of each design is made, so one specimen is poured each day following the same procedure.

The RC beam specimens are tested on a Baldwin 266.9 kN (60 kip) load frame universal testing machine with an ADMET quattro control for the three-point bending load case. A linear variable differential transformer (LVDT) extensometer is used to calculate the deflection at the mid-point. The test is displacement-controlled at a rate of 2mm/minute until failure and the applied force is calculated by the attached load cell.

## 3   Results

The two new tailored TO design approaches are used to design within a prismatic envelope that is 121.9 cm long, 15.2 cm deep, and 7.5 cm wide (48 in x 6 in x 3 in). In both cases, the optimizer is minimizing the system's compliance while limiting the amount of concrete and steel that can be placed. The structures are limited to 9,920 cm$^3$ of concrete (605 in$^3$) and 110 cm$^3$ of steel (6.7 in$^3$), corresponding to a 33% removal of concrete and steel from the prismatic envelope. A Prismatic control beam with the same material volumes as the optimized beams is also designed to have a strength capacity of 13.3 kN (3.0 kips) point load in a 3-point bending load case following the ACI formula for nominal moment capacity [110].

A small amount of post-processing is required to make the optimized design results constructible at the herein used prototype scale. The prototype reinforcement is fabricated by waterjet cutting steel plates. To prepare their fabrication, the truss elements are given a deformation patter to encourage mechanical interlocking between the steel and the concrete [111], [112]. Caps are also added to the ends of the bars following the ACI code 12.1.6 [110]. A small cover layer of concrete is added over the steel, and a mild smoothing operation is used to eliminate sharp edges in the digital model. Detailed drawings of the structural concrete and steel are shown in Fig. 7a, while material volumes after post-processing are shown in Fig. 7b. The structural material volumes are nearly identical in all three designs, with mild variance due to the inclusion and elimination of elements with intermediate densities. The differences in fabricated steel volumes are mainly due to the addition of deformation pattern, which is considered to be non-structural, as the minimum effective area of the reinforcement is not changed. The difference in the concrete volume comes from the addition of the cover layer, which is also assumed to minimally impact structural performance.

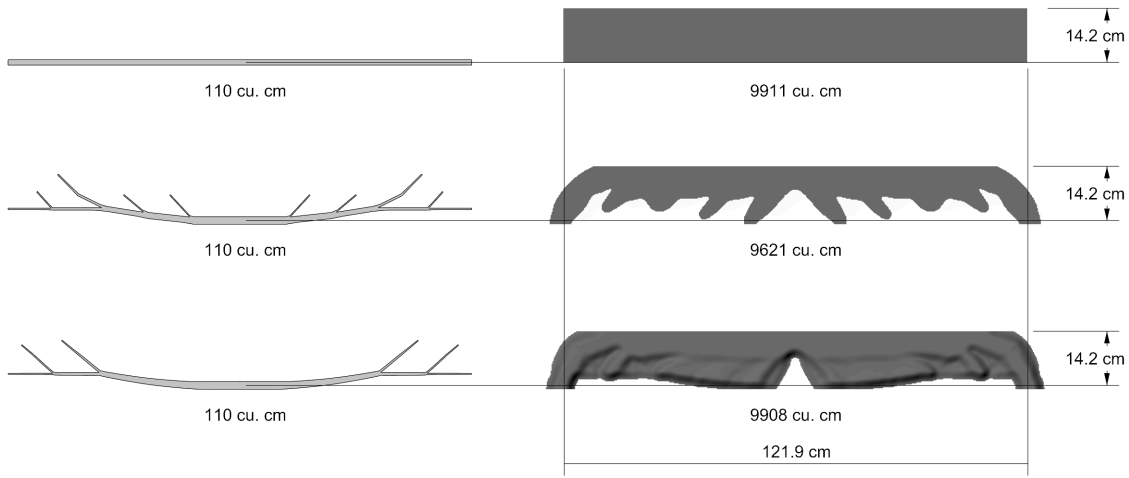

(a)

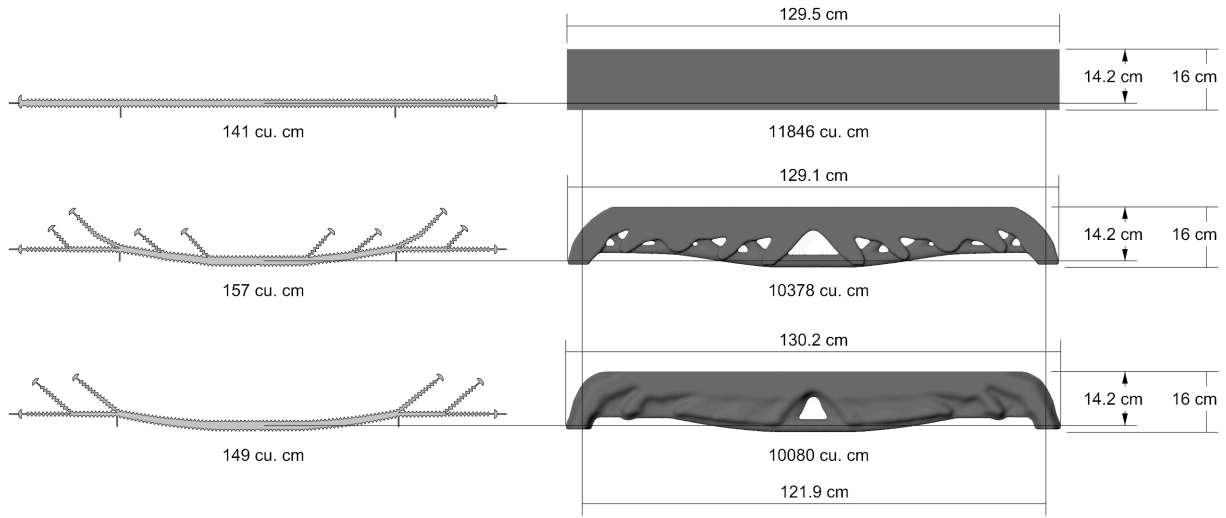

*(b)*

*Figure 7- Structural material of RC beams before and after post-processing. In (a), the design outputs from the optimization algorithms are shown, while (b) shows designs with steel deformations and concrete cover applied. For both (a) and (b), steel elements are on the left, concrete elements on the right, and material volumes are shown beneath each structural component. The Prismatic designs are on the top, the Binary TO in the middle, and the Variable Thickness Sheet TO on the bottom.*

Figure 8 shows the fabricated steel reinforcement for each design when placed into its respective mold, with a photo of the resulting RC beam below. Three specimens of each design are built and experimentally tested. The testing setup is shown in Figure 8d.

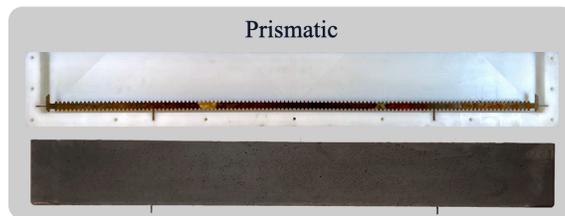

(a)

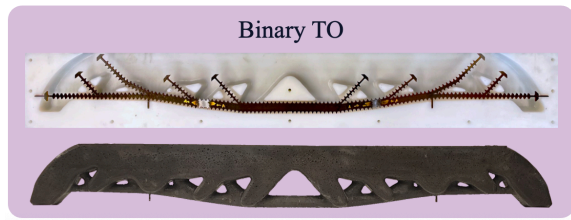

(b)

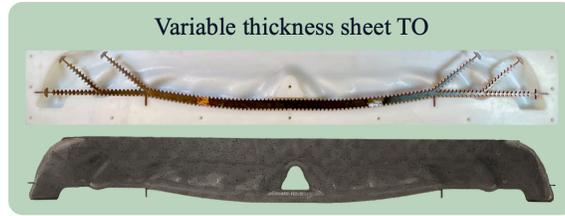

(c)

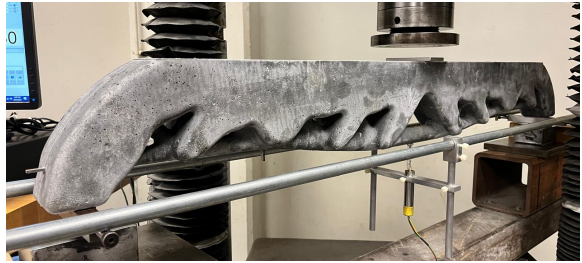

(d)

*Figure 8- Examples of beam fabrication and experimental testing. A picture of the waterjet cut steel reinforcement inserted into a CNC cut mold, and a resulting RC test specimen is shown for (a) a Prismatic sample, (b) a Binary TO sample, and (c) a Variable Thickness Sheet TO sample. A Binary TO beam in the test setup is shown in (d) with an LVDT placed at the beam's center point to measure deflection. The LVDT is mounted directly to the concrete beam supports so that deflection of the steel beam below does not influence the measurements at the midpoint.*

The experimental results are compared in Fig. 9. In Fig. 9a, the load-displacement curves for all tested specimens are shown, with both load and displacement measured at the mid-point. These plots also include three horizontal lines that indicate the average maximum load of the beams before failure, as well as analytical predictions for the design load of prismatic beams with thicknesses of 5.7 cm (2.25 in) and 7.6 cm (3 in), calculated following the ACI code [110]. The analytical prediction for the 5.7 cm (2.25 in) beam gives the design load for the tested Prismatic samples at 13.5 kN. The 7.6 cm (3 in) analytical prediction indicates the expected performance of a Prismatic beam with a higher concrete volume that occupies 100% of the design envelope at 19.2 kN. To ensure this calculation represents an upper bound, it uses the maximum cross-sectional area of steel from the optimized designs of 1.3 cm$^2$ (0.2 in$^2$).

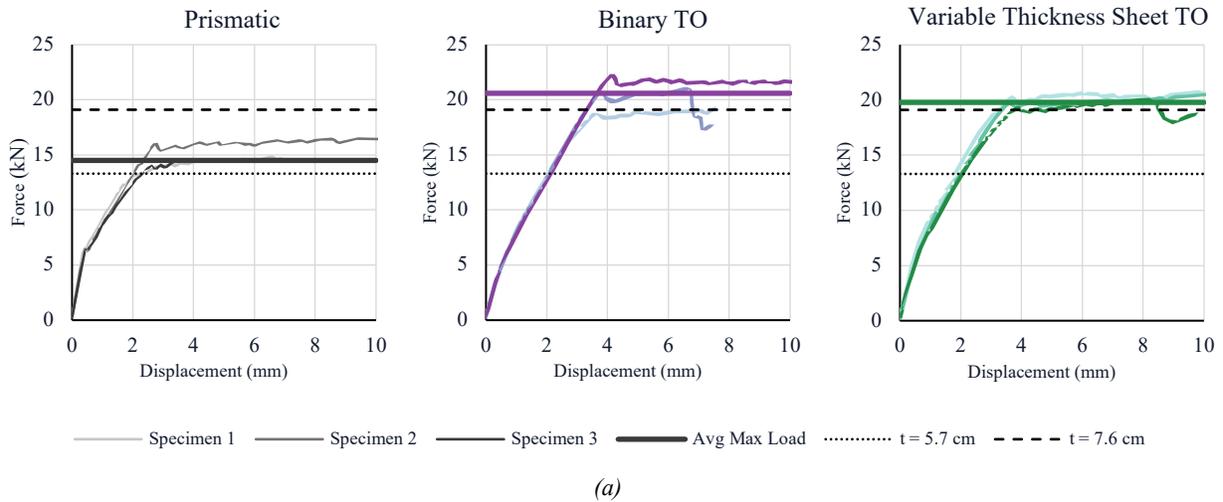

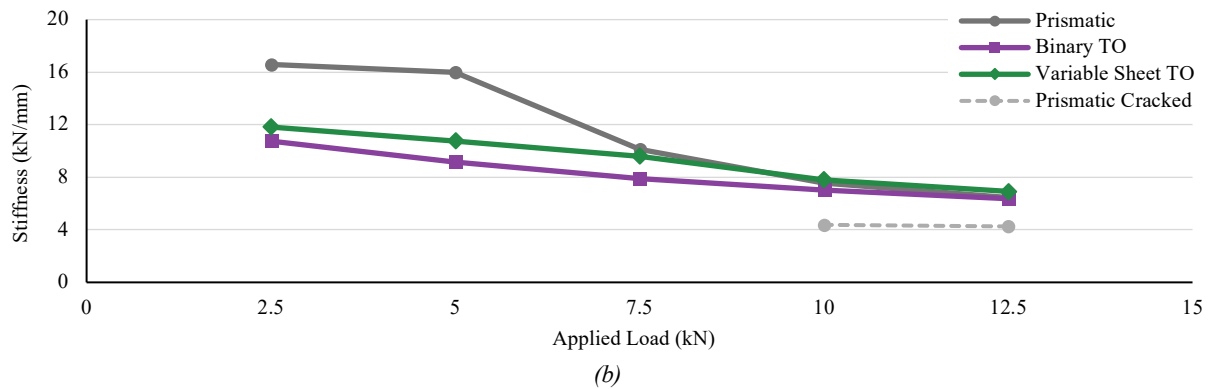

*Figure 9- Experimentally obtained (a) load-displacement curves, and (b) average stiffnesses at different applied load levels. In (a), horizontal lines are added to indicate the average maximum load, and analytical design load predictions for prismatic beams with the same material consumption (66% concrete in the design envelope at t = 5.7 cm) and for designs that fill the design envelope 100% (t=7.6 cm). In (b), the stiffness of the Prismatic samples after cracking is also included.*

All specimens are seen to have a largely linear-elastic behavior until they reach their ultimate load level, after which a near-plastic ductile response becomes evident. All beams exhibit high ductility after failure because the steel yields before the concrete crushes or stress concentrations develop at the joints, despite having significant material removed. A high level of consistency is found in the response of specimens of the same design type, even though the optimized designs sometimes failed in different locations. This suggests the optimized beams are being evenly utilized throughout as an outcome of the optimization process. The load tests were terminated after the deflection exceeded 7 mm, except for the Prismatic Specimen 3, which was terminated prematurely.

From the average maximum load before failure for each design in Fig. 9a, it is seen that the two TO designs fail at higher loads than the Prismatic specimens. The Prismatic design, on average, fails at 14.5 kN, while the Binary TO fails at 20.6 kN, and the Variable Thickness Sheet TO fails at 19.8 kN. This is approximately 42% and 36% higher than the Prismatic design. It should be noted that there is more variance in the failure loads of the Binary TO design. Although maximum load is not the objective function of these new design frameworks, it is a notable emergent behavior. With this in mind, the experimental results show that with the same amount of material as used in conventional prismatic beams, a significant improvement can be achieved in terms of load capacity for the herein considered case study. The results firmly indicate that with the same concrete material consumption, tailored TO algorithms enable RC designs with significantly higher load capacities, without sacrificing ductility.

As mentioned, analytically predicted maximum design loads are also shown in Fig. 9a for prismatic beams with widths of 5.7 cm (2.25 in) and 7.6 cm (3 in). As can be seen, the Prismatic specimens closely approximate the analytical design load. The measured average maximum load is only 9% higher than the analytical prediction. Moreover, it is noteworthy that the TO beams perform nearly as well as a full 7.6 cm beam, which is the equivalent of filling the entire design envelope with material. This suggests that removing material using TO could allow designers to reduce material use by 33% without sacrificing strength performance, and without adding additional depth to the structural elements.

The objective of the TO designs is to maximize the structural stiffness. Therefore, the average stiffness of the three tested beam designs is calculated within the approximate linear range of the experiment at increments of 2.5 kN. These values are shown in Fig. 9b. The Variable Thickness Sheet TO design is seen to have higher stiffness than the Binary TO at all considered loads by approximately 10%. This supports the theory that the VTS method for TO is superior to the Binary optimization for maximizing stiffness [107]. It is also clear that the stiffness of the Prismatic beam specimens exceeds both TO designs at low load values. This is because the concrete is contributing to the stiffness of the beam before cracking occurs. This behavior is not captured by two developed TO algorithms because stiffness values for the materials were specifically chosen so as not to contribute to the performance of the structure in tension. However, as observed, this assumption is not altogether accurate when the beam remains uncracked. It is worth noting that after cracking, which occurs at roughly 6 kN in all Prismatic specimens, the slope of the force-deflection curve drops significantly (Fig. 9b). The effect of this is captured by evaluating the cracked stiffness of the Prismatic beams, which is shown as the dashed line in the stiffness plot in Fig. 9b. These stiffness values are calculated by measuring the slope of the line starting at its deflection when the applied load is 7.5 kN, rather than at 0 kN, as is done for the other stiffness calculations.

# 4 Conclusion

This contribution has demonstrated that when departing from conventional Prismatic shapes, it is possible to achieve 36%-42% higher maximum loads for RC beams without changing the construction material consumption or structural depth. The increased load capacity is accomplished by tailoring free-form structural optimization to the behavioral and fabrication characteristics of RC construction. Specifically, this work has developed and experimentally validated two novel TO algorithms for the design of low-weight RC structures. The algorithms build on previous scholarship by using a hybrid truss-continuum discretization with force-dependent stiffnesses to represent the behavior of steel and concrete. This work expands upon previous TO design for RC by enabling the nodes of the truss elements to move during the optimization to improve their placement and the overall performance of the structure. A traditional Binary TO formulation is compared to a Variable Thickness Sheet TO approach (with a minimum thickness requirement) that can take advantage of concrete's formability. Two TO beam designs are obtained using the new tailored frameworks for RC, and tested to evaluate structural performance.

For both designs, ductile failure is observed, which is critical for RC structures. It is believed that this behavior is achieved through the integration of conventional concrete design principles, such as the reinforcement ratio, minimum feature sizes, and concrete cover layer. The ability to move the truss nodes during the optimization is also a likely contributing factor. However, further experimental tests will be required to understand the most important aspects of optimized RC design.

Through comparison to analytical predictions, the observed potential for material reduction while maintaining today's performance requirements without adding structural depth is around 33%, indicating a viable path forward in reaching carbon neutrality of RC construction, as the current global target reduction requires at least a 22% decrease in concrete consumption [17].

The behavior of all tested TO specimens is relatively consistent, with no clear indications of superior performance of one of the two developed algorithms. However, although the Variable Thickness Sheet TO design has a slightly higher stiffness than the Binary TO result, it poses problems in how the truss and continuum elements interact with one another during the optimization. Further, when imposing the minimum thickness, many of the benefits of VTS are lost. In particular, the problem is no longer convex, and requires parameter tuning. Therefore, in future work, a 2D or 3D TO method is recommended for RC design. Moreover, the stiffness of the Prismatic beam specimens in the linear range is experimentally shown to exceed that of the TO designs, because the used TO discretization ignores the stiffness contribution of plain concrete in tension before cracking. For future work, it is therefore advised that the objective function or material model of the system be modified to more accurately reflect desired performance.

# 5 Acknowledgements

The authors thank Steve Rudolph and Mike Tarkanian for lab assistance in fabrication and testing of the concrete beam specimen. The authors also thank Pitipat "Paul" Wongsittikan and Kittipong "Champ" Thiamchaiboonthawee for their efforts in helping cast and test the specimen, as well as documenting the research. The authors thank Quikrete for providing materials for experiments.

# 6 References

[1] N. Pressmair *et al.*, "Bridging the gap between mathematical optimization and structural engineering: Design, experiments and numerical simulation of optimized concrete girders," *Struct. Concr.*, vol. 24, no. 4, pp. 5314–5330, Aug. 2023, doi: 10.1002/suco.202201096.
[2] J. L. Jewett and J. V. Carstensen, "Experimental investigation of strut-and-tie layouts in deep RC beams designed with hybrid bi-linear topology optimization," *Eng. Struct.*, vol. 197, p. 109322, Oct. 2019, doi: 10.1016/j.engstruct.2019.109322.
[3] G. Gaganelis and P. Mark, "Downsizing weight while upsizing efficiency: An experimental approach to develop optimized ultra-light UHPC hybrid beams," *Struct. Concr.*, vol. 20, no. 6, pp. 1883–1895, Dec. 2019, doi: 10.1002/suco.201900215.
[4] P. Fennell, J. Driver, C. Bataille, and S. J. Davis, "Going net zero for cement and steel," *Nature*, vol. 603, Mar. 2022.

[5] "Cement Statistics and Information | U.S. Geological Survey." Accessed: Oct. 26, 2025. [Online]. Available: https://www.usgs.gov/centers/national-minerals-information-center/cement-statistics-and-information

[6] G. Churkina *et al.*, "Buildings as a global carbon sink," *Nat. Sustain.*, vol. 3, no. 4, pp. 269–276, Jan. 2020, doi: 10.1038/s41893-019-0462-4.

[7] J. Lehne and F. Preston, "Making Concrete Change: Innovation in Low-carbon Cement and Concrete." The Royal Institute of International Affairs Chatham House, 2018.

[8] H. Van Damme, "Concrete material science: Past, present, and future innovations," *Cem. Concr. Res.*, vol. 112, pp. 5–24, Oct. 2018, doi: 10.1016/j.cemconres.2018.05.002.

[9] B. L. Damineli, F. M. Kemeid, P. S. Aguiar, and V. M. John, "Measuring the eco-efficiency of cement use," *Cem. Concr. Compos.*, vol. 32, no. 8, pp. 555–562, Sep. 2010, doi: 10.1016/j.cemconcomp.2010.07.009.

[10] P. J. M. Monteiro, S. A. Miller, and A. Horvath, "Towards sustainable concrete," *Nat. Mater.*, vol. 16, no. 7, pp. 698–699, Jul. 2017, doi: 10.1038/nmat4930.

[11] S. A. Miller, V. M. John, S. A. Pacca, and A. Horvath, "Carbon dioxide reduction potential in the global cement industry by 2050," *Cem. Concr. Res.*, vol. 114, pp. 115–124, Dec. 2018, doi: 10.1016/j.cemconres.2017.08.026.

[12] M. S. Imbabi, C. Carrigan, and S. McKenna, "Trends and developments in green cement and concrete technology," *Int. J. Sustain. Built Environ.*, vol. 1, no. 2, pp. 194–216, Dec. 2012, doi: 10.1016/j.ijsbe.2013.05.001.

[13] "The role of concrete in life cycle greenhouse gas emissions of US buildings and pavements." Accessed: Aug. 14, 2025. [Online]. Available: https://www.pnas.org/doi/epdf/10.1073/pnas.2021936118

[14] K. M. Liew, A. O. Sojobi, and L. W. Zhang, "Green concrete: Prospects and challenges," *Constr. Build. Mater.*, vol. 156, pp. 1063–1095, Dec. 2017, doi: 10.1016/j.conbuildmat.2017.09.008.

[15] K. Scrivener *et al.*, "Impacting factors and properties of limestone calcined clay cements (LC3)," *Green Mater.*, vol. 7, no. 1, pp. 3–14, Dec. 2018, doi: 10.1680/jgrma.18.00029.

[16] A. Naqi and J. G. Jang, "Recent Progress in Green Cement Technology Utilizing Low-Carbon Emission Fuels and Raw Materials: A Review," *Sustainability*, vol. 11, no. 2, 2019, doi: 10.3390/su11020537.

[17] "Roadmap to Carbon Neutrality." American Cement Association, 2021. [Online]. Available: https://www.cement.org/wp-content/uploads/2024/05/Roadmap_Jan2024.pdf

[18] D. Fang, N. Brown, C. De Wolf, and C. Mueller, "Reducing embodied carbon in structural systems: A review of early-stage design strategies," *J. Build. Eng.*, vol. 76, p. 107054, Oct. 2023, doi: 10.1016/j.jobe.2023.107054.

[19] A. Akbarnezhad and J. Xiao, "Estimation and Minimization of Embodied Carbon of Buildings: A Review," *Buildings*, vol. 7, no. 1, 2017, doi: 10.3390/buildings7010005.

[20] F. Pomponi and A. Moncaster, "Embodied carbon mitigation and reduction in the built environment – What does the evidence say?," *J. Environ. Manage.*, vol. 181, pp. 687–700, Oct. 2016, doi: 10.1016/j.jenvman.2016.08.036.

[21] United Nations Environment Programme, "2022 Global Status Report for Buildings and Construction: Towards a Zero-emission, Efficient and Resilient Buildings and Construction Sector," Nov. 2022, [Online]. Available: https://wedocs.unep.org/20.500.11822/41133

[22] E. Hertwich *et al.*, "Resource Efficiency and Climate Change: Material Efficiency Strategies for a Low-Carbon Future," Zenodo, Jan. 2020. doi: 10.5281/ZENODO.3542680.

[23] D. Yeo and R. D. Gabbai, "Sustainable design of reinforced concrete structures through embodied energy optimization," *Energy Build.*, vol. 43, no. 8, pp. 2028–2033, Aug. 2011, doi: 10.1016/j.enbuild.2011.04.014.

[24] M. Afzal, Y. Liu, J. C. P. Cheng, and V. J. L. Gan, "Reinforced concrete structural design optimization: A critical review," *J. Clean. Prod.*, vol. 260, p. 120623, Jul. 2020, doi: 10.1016/j.jclepro.2020.120623.

[25] A. Jayasinghe, J. Orr, T. Ibell, and W. P. Boshoff, "Minimising embodied carbon in reinforced concrete beams," *Eng. Struct.*, vol. 242, p. 112590, Sep. 2021, doi: 10.1016/j.engstruct.2021.112590.

[26] A. Prakash, S. K. Agarwala, and K. K. Singh, "Optimum design of reinforced concrete sections," *Spec. Issue Comput. Eng. Mech.*, vol. 30, no. 4, pp. 1009–1011, Jan. 1988, doi: 10.1016/0045-7949(88)90142-3.

[27] S. Kanagasundaram and B. L. Karihaloo, "Minimum cost design of reinforced concrete structures," *Struct. Optim.*, vol. 2, no. 3, pp. 173–184, Sep. 1990, doi: 10.1007/BF01836566.

[28] V. Govindaraj and J. V. Ramasamy, "Optimum detailed design of reinforced concrete continuous beams using Genetic Algorithms," *Comput. Struct.*, vol. 84, no. 1, pp. 34–48, Dec. 2005, doi: 10.1016/j.compstruc.2005.09.001.

[29] M. Mangal and J. C. P. Cheng, "Automated optimization of steel reinforcement in RC building frames using building information modeling and hybrid genetic algorithm," *Autom. Constr.*, vol. 90, pp. 39–57, Jun. 2018, doi: 10.1016/j.autcon.2018.01.013.

[30] U. Kirsch, "Multilevel Optimal Design of Reinforced Concrete Structures," *Eng. Optim.*, vol. 6, no. 4, pp. 207–212, Jan. 1983, doi: 10.1080/03052158308902471.

[31] J. V. Martí, T. García-Segura, and V. Yepes, "Structural design of precast-prestressed concrete U-beam road bridges based on embodied energy," *J. Clean. Prod.*, vol. 120, pp. 231–240, May 2016, doi: 10.1016/j.jclepro.2016.02.024.

[32] M. A. Ismail and C. T. Mueller, "Minimizing embodied energy of reinforced concrete floor systems in developing countries through shape optimization," *Eng. Struct.*, vol. 246, p. 112955, Nov. 2021, doi: 10.1016/j.engstruct.2021.112955.

[33] M. Nuh, R. Oval, J. Orr, and P. Shepherd, "Digital fabrication of ribbed concrete shells using automated robotic concrete spraying," *Addit. Manuf.*, vol. 59, p. 103159, Nov. 2022, doi: 10.1016/j.addma.2022.103159.

[34] A. Siddika, Md. A. A. Mamun, W. Ferdous, A. K. Saha, and R. Alyousef, "3D-printed concrete: applications, performance, and challenges," *J. Sustain. Cem.-Based Mater.*, vol. 9, no. 3, pp. 127–164, May 2020, doi: 10.1080/21650373.2019.1705199.

[35] E. Lloret *et al.*, "Complex concrete structures: Merging existing casting techniques with digital fabrication," *Mater. Ecol.*, vol. 60, pp. 40–49, Mar. 2015, doi: 10.1016/j.cad.2014.02.011.

[36] J. Garbett, A. P. Darby, and T. J. Ibell, "Optimised Beam Design Using Innovative Fabric-Formed Concrete," *Adv. Struct. Eng.*, vol. 13, no. 5, pp. 849–860, Oct. 2010, doi: 10.1260/1369-4332.13.5.849.

[37] M. Popescu *et al.*, "Structural design, digital fabrication and construction of the cable-net and knitted formwork of the KnitCandela concrete shell," *Structures*, vol. 31, pp. 1287–1299, Jun. 2021, doi: 10.1016/j.istruc.2020.02.013.

[38] A. Jipa and B. Dillenburger, "3D Printed Formwork for Concrete: State-of-the-Art, Opportunities, Challenges, and Applications," *3D Print. Addit. Manuf.*, vol. 9, no. 2, pp. 84–107, Apr. 2022, doi: 10.1089/3dp.2021.0024.

[39] J. Zhang, J. Wang, S. Dong, X. Yu, and B. Han, "A review of the current progress and application of 3D printed concrete," *Compos. Part Appl. Sci. Manuf.*, vol. 125, p. 105533, Oct. 2019, doi: 10.1016/j.compositesa.2019.105533.

[40] W. J. Hawkins *et al.*, "Flexible formwork technologies – a state of the art review," *Struct. Concr.*, vol. 17, no. 6, pp. 911–935, 2016, doi: 10.1002/suco.201600117.

[41] Fardhosseini Mohammad Sadra, Karji Ali, Dossick Carrie Sturts, Lee Hyun Woo, Jebelli Houtan, and Beatty Sean, "The Cost-Effectiveness of Integrating Digital Fabrication for Concrete Formworks," in *Construction Research Congress 2020*, in Proceedings. , 2020, pp. 1077–1086. doi: 10.1061/9780784482865.114.

[42] C. Menna *et al.*, "Opportunities and challenges for structural engineering of digitally fabricated concrete," *Cem. Concr. Res.*, vol. 133, p. 106079, Jul. 2020, doi: 10.1016/j.cemconres.2020.106079.

[43] L. Breseghello, H. Hajikarimian, H. B. Jørgensen, and R. Naboni, "3DLightBeam+. Design, simulation, and testing of carbon-efficient reinforced 3D concrete printed beams," *Eng. Struct.*, vol. 292, p. 116511, Oct. 2023, doi: 10.1016/j.engstruct.2023.116511.

[44] P. Dombernowsky and A. Søndergaard, "Design, Analysis and Realisation of Topology Optimized Concrete Structures," *J. Int. Assoc. Shell Spat. Struct.*, vol. 53, no. 4, 2012.

[45] M. P. Bendsøe and O. Sigmund, *Topology optimization: theory, methods, and applications*, 2. ed., Corrected printing. in Engineering online library. Berlin Heidelberg: Springer, 2004.

[46] N. Stoiber and B. Kromoser, "Topology optimization in concrete construction: a systematic review on numerical and experimental investigations," *Struct. Multidiscip. Optim.*, vol. 64, no. 4, pp. 1725–1749, Oct. 2021, doi: 10.1007/s00158-021-03019-6.

[47] O. Amir, "A topology optimization procedure for reinforced concrete structures," *Comput. Struct.*, vol. 114–115, pp. 46–58, Jan. 2013, doi: 10.1016/j.compstruc.2012.10.011.

[48] M. Bruggi, "Generating strut-and-tie patterns for reinforced concrete structures using topology optimization," *Comput. Struct.*, vol. 87, no. 23–24, pp. 1483–1495, Dec. 2009, doi: 10.1016/j.compstruc.2009.06.003.

[49] Y. Shao, T. Zhao, J. Yan, C. P. Ostertag, and G. H. Paulino, "Improving the Ductility of Concrete Beams Reinforced with Topologically Optimized Steel," *J. Struct. Eng.*, vol. 151, no. 4, p. 04025018, Apr. 2025, doi: 10.1061/JSENDH.STENG-13908.

[50] Gaynor Andrew T., Guest James K., and Moen Cristopher D., "Reinforced Concrete Force Visualization and Design Using Bilinear Truss-Continuum Topology Optimization," *J. Struct. Eng.*, vol. 139, no. 4, pp. 607–618, Apr. 2013, doi: 10.1061/(ASCE)ST.1943-541X.0000692.

[51] J. K. Guest and C. D. Moen, "Reinforced Concrete Design with Topology Optimization," in *Structures Congress 2010*, Orlando, Florida, United States: American Society of Civil Engineers, May 2010, pp. 445–454. doi: 10.1061/41131(370)39.

[52] Y. Xia, M. Langelaar, and M. A. N. Hendriks, "Optimization-based three-dimensional strut-and-tie model generation for reinforced concrete," *Comput.-Aided Civ. Infrastruct. Eng.*, vol. 36, no. 5, pp. 526–543, May 2021, doi: 10.1111/mice.12614.

[53] Y. Xia, M. Langelaar, and M. A. N. Hendriks, "Optimization-based strut-and-tie model generation for reinforced concrete structures under multiple load conditions," *Eng. Struct.*, vol. 266, p. 114501, Sep. 2022, doi: 10.1016/j.engstruct.2022.114501.

[54] O. M. Querin, M. Victoria, and P. Martí, "Topology optimization of truss-like continua with different material properties in tension and compression," *Struct. Multidiscip. Optim.*, vol. 42, no. 1, pp. 25–32, Jul. 2010, doi: 10.1007/s00158-009-0473-2.

[55] G. Gaganelis, D. R. Jantos, P. Mark, and P. Junker, "Tension/compression anisotropy enhanced topology design," *Struct. Multidiscip. Optim.*, vol. 59, no. 6, pp. 2227–2255, Jun. 2019, doi: 10.1007/s00158-018-02189-0.

[56] Y. Li and Y. M. Xie, "Evolutionary topology optimization for structures made of multiple materials with different properties in tension and compression," *Compos. Struct.*, vol. 259, p. 113497, Mar. 2021, doi: 10.1016/j.compstruct.2020.113497.

[57] Y. Luo and Z. Kang, "Topology optimization of continuum structures with Drucker–Prager yield stress constraints," *Comput. Struct.*, vol. 90–91, pp. 65–75, Jan. 2012, doi: 10.1016/j.compstruc.2011.10.008.

[58] M. Bogomolny and O. Amir, "Conceptual design of reinforced concrete structures using topology optimization with elastoplastic material modeling," *Int. J. Numer. Methods Eng.*, vol. 90, no. 13, pp. 1578–1597, Jun. 2012, doi: 10.1002/nme.4253.

[59] O. Amir and O. Sigmund, "Reinforcement layout design for concrete structures based on continuum damage and truss topology optimization," *Struct. Multidiscip. Optim.*, vol. 47, no. 2, pp. 157–174, Feb. 2013, doi: 10.1007/s00158-012-0817-1.

[60] O. Amir and E. Shakour, "Simultaneous shape and topology optimization of prestressed concrete beams," *Struct. Multidiscip. Optim.*, vol. 57, no. 5, pp. 1831–1843, May 2018, doi: 10.1007/s00158-017-1855-5.

[61] Y. Zelickman and O. Amir, "Layout optimization of post-tensioned cables in concrete slabs," *Struct. Multidiscip. Optim.*, vol. 63, no. 4, pp. 1951–1974, Apr. 2021, doi: 10.1007/s00158-020-02790-2.

[62] E. Shakur, A. Shaked, and O. Amir, "Topology and shape optimization of 3D prestressed concrete structures," *Eng. Struct.*, vol. 321, p. 118936, Dec. 2024, doi: 10.1016/j.engstruct.2024.118936.

[63] G. Vantyghem, W. De Corte, E. Shakour, and O. Amir, "3D printing of a post-tensioned concrete girder designed by topology optimization," *Autom. Constr.*, vol. 112, p. 103084, Apr. 2020, doi: 10.1016/j.autcon.2020.103084.

[64] Y. Li, H. Wu, X. Xie, L. Zhang, P. F. Yuan, and Y. M. Xie, "FloatArch: A cable-supported, unreinforced, and re-assemblable 3D-printed concrete structure designed using multi-material topology optimization," *Addit. Manuf.*, vol. 81, p. 104012, Feb. 2024, doi: 10.1016/j.addma.2024.104012.

[65] M. Smarslik and P. Mark, "Hybrid reinforcement design of longitudinal joints for segmental concrete linings," *Struct. Concr.*, vol. 20, no. 6, pp. 1926–1940, Dec. 2019, doi: 10.1002/suco.201900081.

[66] Z. Du, W. Zhang, Y. Zhang, R. Xue, and X. Guo, "Structural topology optimization involving bi-modulus materials with asymmetric properties in tension and compression," *Comput. Mech.*, vol. 63, no. 2, pp. 335–363, Feb. 2019, doi: 10.1007/s00466-018-1597-2.

[67] P. Wei, H. Ma, and M. Y. Wang, "The stiffness spreading method for layout optimization of truss structures," *Struct. Multidiscip. Optim.*, vol. 49, no. 4, pp. 667–682, Apr. 2014, doi: 10.1007/s00158-013-1005-7.

[68] T. Zegard and G. H. Paulino, "Truss layout optimization within a continuum," *Struct. Multidiscip. Optim.*, vol. 48, no. 1, pp. 1–16, Jul. 2013, doi: 10.1007/s00158-013-0895-8.

[69] W. C. Dorn, R. E. Gomory, and H. Grenberg, "Automatic design of optimal structures," 1964.

[70] M. P. Rossow and J. E. Taylor, "A Finite Element Method for the Optimal Design of Variable Thickness Sheets," *AIAA J.*, vol. 11, no. 11, pp. 1566–1569, Nov. 1973, doi: 10.2514/3.50631.

[71] J. Petersson, "On stiffness maximization of variable thickness sheet with unilateral contact," *Q. Appl. Math.*, vol. 54, no. 3, pp. 541–550, 1996, doi: 10.1090/qam/1402408.

[72] J. Petersson and J. Haslinger, "An approximation theory for optimum sheets in unilateral contact," *Q. Appl. Math.*, vol. 56, no. 2, pp. 309–325, 1998, doi: 10.1090/qam/1622499.

[73] F. Golay and P. Seppecher, "Locking materials and the topology of optimal shapes," *Eur. J. Mech. - ASolids*, vol. 20, no. 4, pp. 631–644, Jul. 2001, doi: 10.1016/S0997-7538(01)01146-9.

[74] S. Czarnecki and T. Lewiński, "On minimum compliance problems of thin elastic plates of varying thickness," *Struct. Multidiscip. Optim.*, vol. 48, no. 1, pp. 17–31, Jul. 2013, doi: 10.1007/s00158-013-0893-x.

[75] N. I. Didenko, "Optimal distribution of bending stiffness of an elastic freely supported plate," *Mech. Solids*, vol. 16, no. 1, pp. 147–158, 1981.

[76] S. Pozo, T. Golecki, F. Gomez, J. Carrion, and B. F. Spencer, "Minimum-thickness method for 2.5D topology optimization applied to structural design," *Eng. Struct.*, vol. 286, p. 116065, Jul. 2023, doi: 10.1016/j.engstruct.2023.116065.

[77] T. Yarlagadda, Z. Zhang, L. Jiang, P. Bhargava, and A. Usmani, "Solid isotropic material with thickness penalization – A 2.5D method for structural topology optimization," *Comput. Struct.*, vol. 270, p. 106857, Oct. 2022, doi: 10.1016/j.compstruc.2022.106857.

[78] R. Giele, J. Groen, N. Aage, C. S. Andreasen, and O. Sigmund, "On approaches for avoiding low-stiffness regions in variable thickness sheet and homogenization-based topology optimization," *Struct. Multidiscip. Optim.*, vol. 64, no. 1, pp. 39–52, Jul. 2021, doi: 10.1007/s00158-021-02933-z.

[79] M. S. I. Smith, D. Fang, C. Mueller, and J. V. Carstensen, "Reducing embodied carbon with material optimization in structural engineering practice: Perceived barriers and opportunities," *J. Build. Eng.*, vol. 95, p. 109943, Oct. 2024, doi: 10.1016/j.jobe.2024.109943.

[80] Y. Xia, M. Langelaar, and M. A. N. Hendriks, "A critical evaluation of topology optimization results for strut-and-tie modeling of reinforced concrete," *Comput.-Aided Civ. Infrastruct. Eng.*, vol. 35, no. 8, pp. 850–869, Aug. 2020, doi: 10.1111/mice.12537.

[81] M. Sung and B. Andrawes, "Topology Optimization of Continuous Precast Prestressed Concrete Bridge Girders Using Shape Memory Alloys," *J. Struct. Eng.*, vol. 149, no. 6, p. 04023051, Jun. 2023, doi: 10.1061/JSENDH.STENG-11999.

[82] M. P. Salaimanimagudam, C. R. Suribabu, G. Murali, and S. R. Abid, "Impact Response of Hammerhead Pier Fibrous Concrete Beams Designed with Topology Optimization," *Period. Polytech. Civ. Eng.*, vol. 64, no. 4, Art. no. 4, Sep. 2020, doi: 10.3311/PPci.16664.

[83] Y. Liu, J. L. Jewett, and J. V. Carstensen, "Experimental Investigation of Topology-Optimized Deep Reinforced Concrete Beams with Reduced Concrete Volume," in *Second RILEM International Conference on Concrete and Digital Fabrication*, F. P. Bos, S. S. Lucas, R. J. M. Wolfs, and T. A. M. Salet, Eds., Cham: Springer International Publishing, 2020, pp. 601–611.

[84] B. Wethyavivorn, S. Surit, T. Thanadirek, and P. Wethyavivorn, "Topology Optimization-Based Reinforced Concrete Beams: Design and Experiment," *J. Struct. Eng.*, vol. 148, no. 10, p. 04022154, Oct. 2022, doi: 10.1061/(ASCE)ST.1943-541X.0003465.

[85] N. Pressmair and B. Kromoser, "A contribution to resource-efficient construction: Design flow and experimental investigation of structurally optimised concrete girders," *Eng. Struct.*, vol. 281, p. 115757, Apr. 2023, doi: 10.1016/j.engstruct.2023.115757.

[86] M. P. Bendsøe, "Optimal shape design as a material distribution problem," *Struct. Optim.*, 1989.

[87] E. Andreassen, A. Clausen, M. Schevenels, B. S. Lazarov, and O. Sigmund, "Efficient topology optimization in MATLAB using 88 lines of code," *Struct. Multidiscip. Optim.*, vol. 43, no. 1, pp. 1–16, Jan. 2011, doi: 10.1007/s00158-010-0594-7.

[88] F. Mohebbi, "2D truss FEM program-by Farzad Mohebbi." MATLAB Central File Exchange. [Online]. Available: https://www.mathworks.com/matlabcentral/fileexchange/54011-2d-truss-fem-program-by-farzad-mohebbi

[89] K. Svanberg, "The method of moving asymptotes—a new method for structural optimization," *Int. J. Numer. Methods Eng.*, vol. 24, no. 2, pp. 359–373, Feb. 1987, doi: 10.1002/nme.1620240207.

[90] B. Bourdin, "Filters in topology optimization," *Int. J. Numer. Methods Eng.*, vol. 50, no. 9, pp. 2143–2158, Mar. 2001, doi: 10.1002/nme.116.

[91] J. K. Guest, J. H. Prévost, and T. Belytschko, "Achieving minimum length scale in topology optimization using nodal design variables and projection functions," *Int. J. Numer. Methods Eng.*, vol. 61, no. 2, pp. 238–254, Sep. 2004, doi: 10.1002/nme.1064.

[92] Y. Li, Y. Lai, G. Lu, F. Yan, P. Wei, and Y. M. Xie, "Innovative design of long-span steel–concrete composite bridge using multi-material topology optimization," *Eng. Struct.*, vol. 269, p. 114838, Oct. 2022, doi: 10.1016/j.engstruct.2022.114838.

[93] Y. Li, P. F. Yuan, and Y. M. Xie, "Topology optimization of structures composed of more than two materials with different tensile and compressive properties," *Compos. Struct.*, vol. 306, p. 116609, Feb. 2023, doi: 10.1016/j.compstruct.2022.116609.

[94] K. Cai, "A simple approach to find optimal topology of a continuum with tension-only or compression-only material," *Struct. Multidiscip. Optim.*, vol. 43, no. 6, pp. 827–835, Jun. 2011, doi: 10.1007/s00158-010-0614-7.

[95] M. Smarslik, M. A. Ahrens, and P. Mark, "Toward holistic tension- or compression-biased structural designs using topology optimization," *Eng. Struct.*, vol. 199, p. 109632, Nov. 2019, doi: 10.1016/j.engstruct.2019.109632.

[96] M. Victoria, O. M. Querin, and P. Martí, "Generation of strut-and-tie models by topology design using different material properties in tension and compression," *Struct. Multidiscip. Optim.*, vol. 44, no. 2, pp. 247–258, Aug. 2011, doi: 10.1007/s00158-011-0633-z.

[97] Z. Du, Y. Zhang, W. Zhang, and X. Guo, "A new computational framework for materials with different mechanical responses in tension and compression and its applications," *Int. J. Solids Struct.*, vol. 100–101, pp. 54–73, Dec. 2016, doi: 10.1016/j.ijsolstr.2016.07.009.

[98] Z. Du, G. Zhang, T. Guo, S. Tang, and X. Guo, "Tension-compression asymmetry at finite strains: A theoretical model and exact solutions," *J. Mech. Phys. Solids*, vol. 143, p. 104084, Oct. 2020, doi: 10.1016/j.jmps.2020.104084.

[99] Z. Du *et al.*, "Analysis and optimization of thermoelastic structures with tension–compression asymmetry," *Int. J. Solids Struct.*, vol. 254–255, p. 111897, Nov. 2022, doi: 10.1016/j.ijsolstr.2022.111897.

[100] Z. Du and X. Guo, "Variational principles and the related bounding theorems for bi-modulus materials," *J. Mech. Phys. Solids*, vol. 73, pp. 183–211, Dec. 2014, doi: 10.1016/j.jmps.2014.08.006.

[101] T. Zhao, A. A. Alshannaq, D. W. Scott, and G. H. Paulino, "Strut-and-Tie Models Using Multi-Material and Multi-Volume Topology Optimization: Load Path Approach.," *ACI Struct. J.*, no. 6, 2023.

[102] R. Oviedo, S. Gutiérrez, and H. Santa María, "Experimental evaluation of optimized strut-and-tie models for a dapped beam," *Struct. Concr.*, vol. 17, no. 3, pp. 469–480, Sep. 2016, doi: 10.1002/suco.201500037.

[103] R. Dörrie *et al.*, "Automated force-flow-oriented reinforcement integration for Shotcrete 3D Printing," *Autom. Constr.*, vol. 155, p. 105075, Nov. 2023, doi: 10.1016/j.autcon.2023.105075.

[104] W. Achtziger, "On simultaneous optimization of truss geometry and topology," *Struct. Multidiscip. Optim.*, vol. 33, no. 4–5, pp. 285–304, Feb. 2007, doi: 10.1007/s00158-006-0092-0.

[105] Q. Xia, M. Y. Wang, and T. Shi, "A method for shape and topology optimization of truss-like structure," *Struct. Multidiscip. Optim.*, vol. 47, no. 5, pp. 687–697, May 2013, doi: 10.1007/s00158-012-0844-y.

[106] M. P. Bendsøe and O. Sigmund, "Material interpolation schemes in topology optimization," *Arch. Appl. Mech. Ing. Arch.*, vol. 69, no. 9–10, pp. 635–654, Nov. 1999, doi: 10.1007/s004190050248.

[107] O. Sigmund, N. Aage, and E. Andreassen, "On the (non-)optimality of Michell structures," *Struct. Multidiscip. Optim.*, vol. 54, no. 2, pp. 361–373, Aug. 2016, doi: 10.1007/s00158-016-1420-7.

[108] M. P. Bendsøe and N. Kikuchi, "Generating optimal topologies in structural design using a homogenization method," *Comput. Methods Appl. Mech. Eng.*, vol. 71, no. 2, pp. 197–224, Nov. 1988, doi: 10.1016/0045-7825(88)90086-2.

[109] S. D. Larsen, O. Sigmund, and J. P. Groen, "Optimal truss and frame design from projected homogenization-based topology optimization," *Struct. Multidiscip. Optim.*, vol. 57, no. 4, pp. 1461–1474, Apr. 2018, doi: 10.1007/s00158-018-1948-9.

[110] D. Darwin, C. W. Dolan, and A. H. Nilson, *Design of concrete structures*, vol. 2. McGraw-Hill Education New York, NY, USA:, 2016.

[111] R. Higuchi, J. L. Jewett, and J. V. Carstensen, "Experimental investigation of ribbing pattern effect on the bonding qualities of water jet cut steel reinforcement," *Archit. Struct. Constr.*, vol. 2, no. 3, pp. 455–463, Nov. 2022, doi: 10.1007/s44150-022-00068-3.

[112] M. Takalloozadeh, M. Gilbert, D. Allen, and G. Torelli, "Experimental and numerical assessment of the bond behaviour of laser-cut reinforcement," *Constr. Build. Mater.*, vol. 449, p. 137719, Oct. 2024, doi: 10.1016/j.conbuildmat.2024.137719.